\documentclass[]{article} 

\title{\LARGE \bf
A Game-Theoretical Approach for Finding Optimal Strategies in an Intruder Classification Game\thanks{This work was supported by AFOSR grant FA9550-09-1-0049.}}

\usepackage{graphicx}
\usepackage{color}
\usepackage{multirow}
\usepackage{amsmath,amssymb,amsfonts,amsthm}
\usepackage{dsfont}

\usepackage{balance}

\usepackage{url}
\usepackage{algorithm}
\usepackage{algorithmic}
\usepackage{enumerate}
\usepackage[usenames,dvipsnames]{xcolor}

\graphicspath{{/Users/Lenia/Dropbox/Intruder_classification/}}

\date{}
\newtheorem*{theorem}{Theorem}
\newtheorem{lemma}{Lemma}
\newtheorem{definition}{Definition}
\newcommand{\vect}[1]{\boldsymbol{#1}}
\newcommand{\eat}[1]{}

\author{Lemonia Dritsoula\thanks{Lemonia Dritsoula and John Musacchio are with University of California, Santa Cruz, CA, USA.
        {\tt\small \{lenia, johnm\}@soe.ucsc.edu}}, Patrick Loiseau \thanks{Patrick Loiseau is with EURECOM, Sophia-Antipolis, France. Part of this work was done while he was at UC Santa Cruz, CA, USA. {\tt\small patrick.loiseau@eurecom.fr}} , and John Musacchio\footnotemark[2]}%

\begin{document}
\maketitle
\thispagestyle{empty}
\pagestyle{empty}
\begin{abstract}

We consider a game in which a strategic defender classifies
an intruder as spy or spammer. The classification is based on the
number of file server and mail server attacks observed during a fixed
window. The spammer naively attacks (with a known distribution) his
main target: the mail server. The spy strategically selects the number
of attacks on his main target: the file server. The defender
strategically selects his classification policy: a threshold on the
number of file server attacks. We model the interaction of the two
players (spy and defender) as a nonzero-sum game: 
The defender needs to balance missed detections and false alarms in 
his objective function, while the spy has a tradeoff between attacking
 the file server more aggressively and increasing the chances of getting caught.
We give a characterization of the Nash equilibria in mixed strategies,
and demonstrate how the Nash equilibria can be computed in polynomial
time. Our characterization gives interesting and non-intuitive insights
on the players' strategies at equilibrium: The defender uniformly randomizes
between a set of thresholds that includes very large values. The strategy of the spy 
is a truncated version of the spammer's distribution. We
present numerical simulations that validate and illustrate our theoretical results.
\end{abstract}

\section{Introduction}

Cybersecurity is important to businesses and individuals. According to
a recent study conducted by Symantec~\cite{Symantec}, the number of
cyber attacks and threats has increased during 2011,
resulting in lost productivity, reduced revenue, and bad reputation for the
associated businesses. Different kinds of attacks (e.g., internal
unintentional actions and external malicious ones) should be
treated differently and organizations need the security intelligence
to respond to all threats rapidly. Since only less than half of
organizations are currently pursuing security issues, there is 
still room for improvement. Our work contributes to the understanding of
the interaction between network operators and potential attackers.
 
In almost every network security situation, the administrator of a
network (defender) has limited resources in energy and
time. The defender needs to distinguish between different types of
attackers (spy or spammer) and decide whether to take actions or
not. For example, an attack on a mail server by a spammer (causing at
most network congestion) should be treated differently than an attack
on a file server (possibly involving identity theft). Therefore, the defender
should employ various statistical tests, based on the observed
number of file and mail server attacks and decide upon the true
type of the attacker. Knowing that a defender is trying to classify
attackers, the strategic spy is likely to change the way he attacks in
order to make it more difficult to be classified as a spy. In this
work, we analyze a simple model of such a classification game and
extract key insights.

There exists a growing body of works on the topic of intrusion
detection. In a series of papers~\cite{AlB03,AlB04, AlB06}, Alpcan and Ba\c{s}ar
present a game-theoretic analysis of a security game between an attacker and
an intrusion detection system in different scenarios, both in finite and
continuous-kernel versions. Our
game-theoretic framework focuses on attacker classification, rather than intrusion
detection. In the presence of a non-strategic player who is represented with a fixed and
known probability distribution, the defender's task of distinguishing the true type of 
the attacker becomes more challenging. It is also interesting to see how the nonstrategic spammer influences the spy's strategy. 

In~\cite{Patcha}, Patcha and Park model
the interaction between a node who might be regular or dangerous and a defender
who has some prior probability for the existence of each type in an ad hoc network. They consider a signaling
and dynamic game with multiple stages and the players update their beliefs and distributions 
based on Bayes' rule. On the contrary, our work considers a one-stage game with a fixed duration
 and we compute the Nash equilibria in mixed strategies.

Chen and Leneutre~\cite{Leneutre} address the intrusion detection
problem in heterogeneous networks consisting of nodes with different
non-correlated security assets, in the same way that the file and mail servers are
of different importance in our work. They consider a static game with
full information and limited attack and monitoring resources. We do not consider such limitations
and we assume asymmetric information, since the defender is not aware of the
attacker's type.

Bao, Kreidl, and Musacchio~\cite{BKM2011} also consider an intruder classification
game, in which the sequence of attacks is taken into account. While their model has many similarities with ours, 
we focus on less complex (but still realistic) payoff functions that allow us to go 
one step further than simulations and analyze the structure of the Nash equilibria.

Gueye, Walrand, and Anantharam~\cite{assGMNT11,ass} have investigated the structure of the Nash equilibria in a network topology game where two adversaries select which links to attack and defend. They consider a special case of nonzero-sum games, in which the different term in the players' payoffs is controlled only by the one player. In these games,  one player optimizes his payoff against the other who has optimized his payoff as well. Such games are easier to analyze than general nonzero-sum games, and they give interesting insights on the strategies of the two players. Our work is using a similar payoff formulation in a different setting: the defender selects a threshold on file server attacks (not a set of links to defend) and there are two different types of attackers.

To the best of our knowledge,~\cite{KDD04} is the most relevant work to ours. The authors address the problem of classifying a malicious intruder in the presence of an innocent user. The malicious intruder can perturb his behavior to confuse the classifier and evade detection. But, their work focuses only on one iteration of the game and how each player can once adjust his strategy to optimize his expected payoff, given the optimal strategy of the other player. On the contrary, we provide an algorithm to find the Nash equilibria of the game. 

In summary, our contributions are the following. We propose a game-theoretic model to analyze the interactions
  between two adversaries: a classifier (defender) and a malicious attacker when a nonstrategic 
  spammer is present. We compute the Nash
  equilibria in polynomial time. We perform numerical experiments that validate the theoretical computation and give non-intuitive insights on the players' strategies.

The rest of the paper is organized as follows. Section~\ref{game_model} describes
the game model and underlying assumptions. Section~\ref{analysis} explains how to reduce the complexity of the
game and compute the Nash equilibria in polynomial time. Section~\ref{sims}
presents the performance evaluation through numerical experiments and
Section~\ref{concl} concludes the paper.


\section{Game Model}
\label{game_model}

The game model is as follows. A network consists of a defender and two servers that are monitored for potential attacks: a File Server (FS) with sensitive data and a Mail Server (MS) with contents of inferior importance. We assume a constant classification window of $N$ time slots (discrete time), during which the defender observes the number of hits on the FS / MS coming from a single attacker. Nature decides the type of the attacker in the network: spy or spammer with probabilities $p$ and $1-p$ respectively. 

The defender is a strategic player and seeks to correctly classify the potential intruder. He selects a threshold $T$. If he observes $T$ or more hits on the FS, he classifies the attacker as a spy; otherwise as a spammer. The spy's goal is to attack the FS as frequently as possible while evading detection. He is a strategic player and selects the number of FS attacks $H$ to launch. 

The spammer is a non-strategic player with a tendency to attack more often the MS to congest the network or annoy the defender. He attacks the FS $Z$ time slots with a known and fixed distribution. For instance, he can be modeled to have a Bernoulli distribution at each time slot with a small per-period probability $\theta_0$ of a hit on the FS.

\vspace{0.1in}
\textbf{Notational Conventions:}\\
We use ``$\min[\vect v]$" to denote the minimum element of
  a vector $\vect v$, and ``$\text{minimize}$" when we minimize a specific
  expression over some constraints. We use the \emph{prime} sign ($'$) for transpose of matrices and vectors. All vectors are assumed to be column vectors and are denoted by bold lowercase letters (e.g., $\vect \alpha$, $\vect \beta$). For matrix notation we use capital greek letters (e.g., $\Lambda$) and 
the indicator function is denoted by $\mathds{1}_{cond}$. It is equal to 1 if ``cond"  holds and is equal to 0 otherwise. 
The column vector of ones of length $N$ is denoted by $\vect 1_{N}$ and the matrix of ones of dimensions $N \times M$ is denoted by $1_{N \times M}$. An overview of the important parameters is shown in Table~\ref{references2}.

\subsection{Spy's cost function}

The spy is detected when $T \leq H$, which incurs a cost of $c_d$ to the spy. Each of the $H$ FS hits gives the spy a benefit of $c_a$. We assume that the spy gains nothing from attacking the MS. We will work with a cost function for the attacker rather than a payoff function, thus, his overall cost function can be expressed as follows
\begin{equation*}
  J_A(T,H) = c_d \cdot \mathds{1}_ {T \leq H} - c_a \cdot H. 
\end{equation*}
\eat{where $\mathds{1}_{T \leq H}$ is $1$ if  $T \leq H$ and $0$ otherwise.}

\subsection{Defender's reward function}

The defender's expected reward function depends on the true type of the attacker. 

\begin{itemize}
\item With probability $p$ the defender faces a spy and classifies him correctly when $T \leq H$. The defender gains $c_d$ for the correct classification of the spy, but loses $c_a$ per FS hit.  
\item With probability $1-p$ the defender faces a spammer, who is incorrectly classified as spy with probability $\phi (T) = \Pr\{Z \geq T\}$. The expected false alarm penalty in this case is $c_{fa} \cdot \phi(T)$. 
\end{itemize}
Combining these two scenarios, the defender's expected payoff is 
\begin{equation*}
  \tilde{U}_D(T,H) = p \cdot \left(c_d \cdot \mathds{1}_ {T \leq H} - c_a \cdot H\right) - (1-p) \cdot c_{fa} \cdot
  \phi(T).
\end{equation*}
By scaling the above function, we get 
\begin{equation*}
  U_D(T,H) = c_d \cdot \mathds{1}_ {T \leq H}  - c_a \cdot H - \mu(T),
\end{equation*}
where $\mu(T)= \displaystyle \frac{1-p}{p} \cdot c_{fa} \cdot \phi(T).$

Function $\phi(T)$ is decreasing (since it is an complementary cumulative density function) and we also assume that it is strictly decreasing: $\Pr\{Z \geq T\} > \Pr\{Z \geq T+1\}$. 

\subsection{Players' interactions}
For a fixed observation window $N$ the spy has $N+1$ available
actions (attack the file server $H \in \{0,\ldots,N\}$ times), whereas the
defender has $N+2$ available actions (select $T \in \{0,\ldots, N+1\}$ as
the classification threshold). A threshold of $0$ always results in
spy classification (any intruder will attack the FS at least $0$
times); a threshold of $N+1$ always results in spammer classification (a spy cannot attack $N+1$ times during $N$).

We model our problem as a nonzero-sum game. However, the defender's payoff is different from the spy's cost function in only one term $\mu(T)$ that depends only on the defenderÕs strategy ($UD(T,H) = JA(T,H) - \mu(T )$). These games are known as almost zero-sum games or quasi zero-sum games.
We are interested in Nash equilibria in mixed strategies for the
following reason. The spy seeks to select $H$ just below $T$ to evade detection. The
defender aims to select a threshold $T$ equal to the attacker's
strategy $H$. Thus the players need to mix between different strategies to
make themselves less predictable. The spy  chooses a distribution
$\vect{\alpha}$ on the available numbers of FS hits --- thus $\vect{\alpha}$ is a vector of size $N+1$ with non negative elements that sum to 1.
Similarly the defender chooses a distribution $\vect{\beta}$ on the collection of possible thresholds $T$. Thus $\vect{\beta}$ is a vector of size $N+2$.

Let $\tilde{\Lambda}$ be a $(N+1) \times (N+2)$
matrix representing the spy's strategies' cost. Since the number of strategies available to each player is not the same, the cost matrix $\tilde \Lambda$ is not square. We express the cost matrix of the attacker as
\eat{
\[
 \tilde \Lambda =
c_d \cdot \underbrace{ \begin{pmatrix}
  1 & 0 & \cdot \cdot \cdot & \cdot \cdot \cdot & 0  & 0\\
    \vdots & 1 & \ddots &  & \vdots &\vdots \\
    \vdots & & \ddots & \ddots &  \vdots & \vdots \\
    \vdots & && \ddots & 0  &\vdots \\
       1 &\cdots &\cdots& \ldots & 1 &0  \\
 \end{pmatrix}}_{\tilde \Lambda_1} - c_a \cdot  \underbrace{\begin{pmatrix}
   0 & 0  & \cdot \cdot \cdot & 0  \\
    1 & 1  & \cdot \cdot \cdot & 1 \\
    2 & 2  & \cdot \cdot \cdot  & 2 \\
    \vdots  & \vdots  & & \vdots \\
    N-1 & N-1  & \cdot \cdot \cdot & N-1 \\
    N & N & \cdot \cdot \cdot  & N  \\
 \end{pmatrix}}_{\tilde \Lambda_2}
\]
}
\[
 \tilde \Lambda =
c_d \cdot \underbrace{ \begin{pmatrix}
  1 & 0 & \cdot \cdot \cdot & \cdot \cdot \cdot & 0  & 0\\
    \vdots & 1 & \ddots &  & \vdots &\vdots \\
    \vdots & & \ddots & \ddots &  \vdots & \vdots \\
    \vdots & && \ddots & 0  &\vdots \\
       1 &\cdots &\cdots& \ldots & 1 &0  \\
 \end{pmatrix}}_{\tilde \Lambda_1} 
 - c_a \cdot  \underbrace{\begin{pmatrix}
   0   \\
    1  \\
    2  \\
    \vdots  \\
    N-1  \\
    N   \\
 \end{pmatrix} \cdot \vect 1'_{N+2}}_{\tilde \Lambda_2}
 \]with respective elements $\tilde \Lambda_{1}(i,j) =  \mathds{1}_{j \leq i}$ and $\tilde \Lambda_2(i,j) = i$, where $i = \{0, \dotsc, N\}$ designates the row and $j = \{0, \dotsc, N+1\}$ the column.

Each row $i$ of $\tilde \Lambda$ corresponds to one of the $N+1$ possible spy strategies. For instance, row ``0" corresponds to spy attacking the FS 0 times (or $H=0$), row ``1" corresponds to spy selecting $H = 1$ and so on. Each column of  $\tilde \Lambda$ corresponds to one of the $N+2$ possible defender strategies. For instance, column ``0" corresponds to defender selecting $T = 0$ (or always classify as spy). In this case the spy is always detected and loses $c_d$. The last column $``N+1"$ corresponds to defender selecting $T = N+1$. Since it is not possible that that spy attacks $N+1$ times during $N$ time slots, if the defender selects this strategy, the spy is never caught and has zero detection cost.
In $\tilde \Lambda_2$, every column $j$ (defender strategy) incurs the same benefit to the spy. No matter what the decision threshold is, the impact of the spy's attacks is the same.

Let $\tilde{\Lambda}$ be defined as above, and $\vect \alpha$, $\vect \beta$, be the spy and defender distributions respectively. The attacker cost can be written as  $\vect{\alpha}'\tilde{\Lambda} \vect{\beta}$  and the defender payoff can be written as  $\vect{\alpha}'\tilde{\Lambda} \vect{\beta} -  \vect{\mu}'\vect{\beta}$, where $\vect \mu$ is a strictly decreasing vector (component-wise) with $\mu_i$ be the $i^\text{th}$ component of vector $\vect \mu$. Certain computations are simplified by using a matrix with only positive entries. We define
\[ \Lambda  = \tilde{\Lambda} +  (N \cdot c_a + \epsilon)\cdot 1_{(N+1)\times(N+2)}, \]
where $1_{(N+1)\times(N+2)}$ is a matrix of all ones of dimension $(N+1) \times (N+2)$ and $\epsilon > 0$.\eat{, with $\epsilon \rightarrow 0$}
Since $\vect{\alpha}$ and $\vect{\beta}$ must each sum to 1, the expressions $\vect{\alpha}' \Lambda \vect{\beta} $ and $\vect{\alpha}' \Lambda \vect{\beta} -  \vect{\mu}'\vect{\beta}$ are respectively the attacker cost and defender payoff shifted by a constant. Adding a constant to the players' payoff does not affect their best responses, thus from here on we will consider these expressions to be the payoff functions of each player.

\begin{table}[t]
    \caption{Main Notations}
    
    \centering
    \begin{tabular}{|c | c || c | c | }
      \hline
       $p $ & probability for spy &  $\vect \alpha$ & spy's mixed strategy\\
      \hline
      $c_d$ & detection cost & $\vect \beta$ & def. mixed strategy\\
      \hline
     $c_a $ & FS attack cost &   $\vect \mu$ & false alarm cost vector \\
      \hline
      $c_{fa} $ & false alarm penalty  &   $\theta(\vect \beta)$ & defendability of $\vect \beta$\\
      \hline
         $H $ & spy's strategy (\# FS hits) & $\Lambda$ & cost matrix of spy\\
	\hline
         $T $ & def. strategy (threshold) &     $s $ & first tight inequality\\
          \hline
         $Z $ & \# of FS hits by spammer &    $f $ & last tight inequality\\
	\hline     
           \end{tabular}
    \label{references2}
  \end{table}
  \label{references}


\section{Game-Theoretic Analysis}
\label{analysis}

Nash proved in \cite{Nash51} that every finite game (finite number of players with finite number of actions for each player) has a mixed-strategy Nash equilibrium. Our game is finite, thus it admits a NE in mixed strategies. In a two-player game, the players' strategies $\vect \alpha$ and $\vect{\beta}$ are a NE if each player's strategy is a best response to the other player's mixed strategy.

\subsection{Best response analysis}

We will first prove a series of lemmata that will help us state and prove our main Theorem.
We first prove that in a NE, the spy's strategy $\vect{\alpha}$ minimizes his cost and the defender's strategy $\vect{\beta}$ maximizes his payoff. 

\begin{lemma}
A spy who plays a best response to a defender strategy $\vect{\beta}$, has a cost $\delta =  \min [\Lambda \vect \beta]$.
\label{attacker_LP}
\end{lemma}

\begin{proof}
For a given defender strategy $\vect \beta$ and since $\Lambda$ is positive, the minimum attacker cost is achieved by putting positive probability only on strategies corresponding to the minimum entries of the vector $\Lambda \vect \beta$. Thus the spy's optimal cost is $\delta = \min[\Lambda \vect \beta$].
\end{proof}
\begin{definition}[Defendability]
 The defendability of a mixed strategy $\vect \beta$ is defined as

   \begin{equation}
  	\theta (\vect \beta) = \min [{\Lambda \vect \beta] - \vect \mu' \vect \beta}
\label{def_defendability}. 
\end{equation}
It corresponds to the defender's payoff when the attacker plays a best response to $\vect \beta$. 
\end{definition}
The defendability is similar to the notion of vulnerability in \cite{assGMNT11}, that is a measure of how vulnerable a set of links is. An interesting property of the defendability is that it depends only on the defender's strategy and not on the spy's selection of $\vect \alpha$. This is due to the aforementioned ``almost" zero-sum game. We will exploit this property in the subsequent analysis. 

In Nash equilibrium, each player in the game selects a best response to the other player's strategies. We show below (Lemma~\ref{defend_max}) that the defender's best response to any spy's best response maximizes the defendability.

We proved in Lemma~\ref{attacker_LP} that the best response of the spy against any defender strategy $\vect \beta$, gives a spy cost $\delta = \min[\Lambda \vect \beta ]$. 
The attacker's optimization problem, subject to the constraint that he limits the defender to the defendability $\theta(\vect \beta)$ takes the following form 

\vspace{0.1in}
\noindent \textit{\textbf{Primal with constraints:}}
\begin{equation}
  \begin{aligned}
    & \underset{\vect \alpha}{\text{minimize}}
    & &  \vect \alpha' \Lambda \vect \beta\\
    & \text{subject to}
    & & \vect \alpha \geq \vect 0,  \vect 1'_{N+1} \cdot \vect \alpha \geq 1, \\
    & & &\vect \alpha' \Lambda - \vect \mu'  \leq \theta(\vect \beta) \cdot \vect 1'_{N+2}.\\
  \end{aligned}
  \label{min_constrained}
\end{equation}
The last constraint in the above LP comes from the fact that the defender's payoff from any pure strategy in the support of the defender's NE strategy is the same (and equal to the defendability), and at least as good as the payoff from any pure strategy not in the support of his mixed strategy, when the attacker is playing his NE strategy. The dual constrained LP is 
\vspace{0.1in}

\noindent \textit{\textbf{Dual with constraints:}}
\begin{equation}
  \begin{aligned}
    & \underset{\vect y, z}{\text{maximize}}
    & &   (-\vect 1'_{N+2}  \cdot  \theta(\vect{\beta}) - \vect \mu') \vect y  + z\\
    & \text{subject to}
    &&  \vect y \geq \vect 0,~ z \geq 0\\
    & & & z \cdot \vect 1_{N+1} - \Lambda \vect y \leq \Lambda \vect \beta.\\
  \end{aligned}
  \label{opt_dual2}
\end{equation}
As we show below, the optimal value of the dual LP given by (\ref{opt_dual2}) is equal to $\delta$ if and only if $\vect \beta$ is a maximizer of the function $\theta(\vect \beta)$. If the optimal value was greater than $\delta$, then the attacker would not play a best response against a strategy $\vect \beta$, namely we would not be in NE.

Working on (\ref{opt_dual2}), the last constraint
gives $\Lambda (\vect \beta+\vect y) \geq z \cdot \vect 1_{N+1}$, 
and since we seek to maximize a nonnegative $z$ with an upper limit, $z = \min[\Lambda (\vect \beta + \vect y)]$. We note here that since $\Lambda$ and $\left( \vect \beta + \vect y\right)$ are nonnegative matrix and vector respectively, their multiplication is also a nonnegative vector and the above optimal value for $z$ is valid. With the above substitution for $z$, we get the following LP

\begin{equation}
  \begin{aligned}
    & \underset{\vect y}{\text{maximize}}
    & &- \|\vect y\| \theta(\vect \beta) - \vect \mu' \vect y + \min[ {\Lambda(\vect{\beta}+ \vect{y})]}\\ 
    & \text{subject to}
    & &  ~\;\;\vect{y} \geq \vect 0,\\
  \end{aligned}
  \label{last_opt}
\end{equation}
where we define $\|\vect y\| =  \|\vect y\|_1  = \displaystyle\sum_{i=0}^{N+1} |y_i|.$


\begin{lemma}
  In NE, the defender strategy  $\vect{\beta}$ maximizes the defendability $\theta(\vect \beta)$.
  \label{defend_max}
\end{lemma}

\begin{proof}

Part I: Suppose that the defender's strategy is $\vect \beta$ such that $\theta(\vect \beta) < \theta(\vect \xi)$, where $\vect{\xi} = \arg \max \theta$. Let $\vect{y} = k\vect{\xi}$, with $k \gg 1$.
 Then (\ref{last_opt}) gives
  \begin{equation}
  \begin{aligned}
    & \underset{k}{\text{maximize}}
    & & - \|k \vect \xi \| \theta(\vect \beta) - \vect \mu' k \vect \xi +
  \min[\Lambda(\vect \beta+k\vect \xi)]\\
    & \text{subject to}
    & &  ~\;\;k \gg 1.\\
  \end{aligned}
   \label{befgreater2}
\end{equation}
Since $k \vect \xi \gg \vect \beta$, $\|\vect \xi \| = 1$ and
$\theta(\vect \xi) = \min[\Lambda \vect \xi] - \vect{\mu}' \vect
\xi$, the argument that needs to be maximized in (\ref{befgreater2}) becomes $
  -k \theta(\vect \beta) + k \theta(\vect \xi)$ or $k \left( \theta(\vect
  \xi) - \theta(\vect \beta)\right)$.
Since $\theta(\vect \xi) > \theta(\vect \beta)$, this expression can be made arbitrarily large. 
Therefore, the optimal value of (\ref{last_opt}) is infinity. Since the optimal value of the dual problem is unbounded, the initial primal problem is
infeasible~\cite{Lue84}. Therefore, if the defendability of $\vect \beta$ is not maximal, then $\vect \beta$ is not a NE.\eat{ Hence, \emph{if the defendability is not maximized, the attacker cannot simultaneously select an optimal strategy and make the defender unable to improve by deviating.}}
  
Part II: Suppose that the defender's strategy $\vect{\beta} \in \arg \max \theta(\vect \lambda)$. We show that the optimal values of the attacker's constrained and unconstrained LP problem are the same, i.e., $\delta_{const} =\delta$, where $\delta_{const}$ is the optimal value of (\ref{min_constrained}). Since (\ref{min_constrained}) is a minimization problem over a smaller set (extra constraints), 
 \begin{equation}
 \delta_{const} \geq \delta.
 \label{condition1}
 \end{equation}
With the change of variable  $\vect q = \vect \beta+ \vect y$, we transform (\ref{last_opt}) to the following problem
\begin{equation}
  \begin{aligned}
    & \underset{\vect q}{\text{maximize}}
    & &- (\|\vect q\| - \|\vect \beta\| )\theta(\vect \beta) - \vect \mu' (\vect q - \vect \beta)+ \min[\Lambda\vect q]\\ 
    & \text{subject to}
    & &  ~\;\;\vect{q} \geq \vect \beta.\\
  \end{aligned}
  \label{last_opt2}
\end{equation}
We take now a relaxed version of the above problem, where the constraint is $\vect q \geq  \vect 0$ instead of $\vect q \geq \vect \beta$
\begin{equation*}
  \underset{\vect{q} \geq \vect 0} {\text{maximize}}\{-(\|\vect q\| - \|\vect \beta\| )
  \theta(\vect \beta) - \vect \mu'(\vect q - \vect \beta) + \min[ \Lambda \vect q]\}.
  \label{relaxed3}
  \end{equation*}
 Clearly the optimal value $\delta_{relax-const}$ in the above relaxed maximization problem is greater than or equal to the optimal value $\delta_{const}$ of the original problem (\ref{last_opt2}), since we maximize the same objective function over a larger set. Thus 
   \begin{equation}\delta_{relax-const} \geq \delta_{const}. 
\label{relaxed1}\end{equation} 
 Since $ \| \vect \beta \| = 1$ and $\delta = \theta(\vect \beta) + \vect \mu' \vect \beta$, from the above relaxed problem we get
 \begin{equation}
  \underset{\vect{q} \geq \vect 0} {\text{maximize} }\{ \delta -\|\vect q\| \theta(\vect
  \beta) +\min[\Lambda \vect q] - \vect \mu' \vect q\}.
  \label{last}
\end{equation}
But $\min[\Lambda \vect q] - \vect \mu' \vect
q = \| \vect q \| \cdot \theta(\vect q) \leq  \| \vect q \| \cdot \theta(\vect \beta),$ since $\vect \beta \in
\arg \max \theta(\lambda)$.  Thus, the maximization in (\ref{last}) always
gives an optimal value \begin{equation} \delta_{relax-const}\leq \delta. \label{relaxed2} \end{equation} 
From (\ref{condition1}), (\ref{relaxed1}) and (\ref{relaxed2}) we get $\delta \leq \delta_{const} \leq \delta_{relax-const} \leq \delta$,
which yields $\delta_{const} = \delta$.
\end{proof}


\begin{definition}[Tight constraint]
 An inequality constraint is tight, if it holds as an equality;
  otherwise, it is said to be loose.
\end{definition}

\begin{definition}[Polyhedron]
A polyhedron is the solution set of a finite number of linear equalities and inequalities.
\end{definition}

\begin{definition}[Extreme point]
  A point $\vect x$ of a polyhedron is said to be extreme if there
  is no $\vect x'$ whose set of tight constraints is a strict superset of the set of tight
  constraints of $\vect x$. 
  \label{extreme}
\end{definition}

\begin{lemma}
There exists a defender NE strategy $\vect \beta$ amongst the extreme points of a polyhedron defined by $\Lambda \vect x \geq \vect {1}_{N+1}$, $\vect x \geq \vect 0$.
\label{def_lemma}
\end{lemma}

\begin{proof}
As we proved in Lemma~\ref{defend_max}, in NE, the defender maximizes the defendability, that is, 
he solves the following ``defendability LP"

\begin{equation}
  \begin{aligned}
    & \underset{\vect \beta, z}{\text{maximize}}
    & &   - \vect \mu' \vect \beta + z\\
    & \text{subject to}
    & & z \cdot \vect 1_{N+1} \leq \Lambda \vect \beta\\
    & & & \vect 1'_{N+2} \cdot \vect{\beta} = 1, ~\vect \beta \geq \vect 0\\ 
  \end{aligned}
  \label{def_LP}
\end{equation}
The solution for $z$ is $z = \min[\Lambda \vect \beta]$ (finite and positive since $\Lambda$ positive). The objective is a finite quantity reduced by a positive value ($\vect \mu \geq \vect 0,  \vect \beta \geq \vect 0$), thus (\ref{def_LP}) is bounded.
Consider a vector $[\vect{\beta} ; z]$ that is both a solution to (\ref{def_LP}) and extreme for the inequalities of (\ref{def_LP}). From the basic theorem of linear programming such a point must exist. Let $S$ be the set of indices of tight inequalities of $\Lambda \vect \beta \geq z \cdot \mathbf{1}_{N+1}$, and $P$ be the set of indices of tight inequalities in $\vect \beta \geq \vect 0$. 

Defining $\vect x:= \vect \beta/ z$, the above constraints become $\Lambda \vect x \geq \vect 1$, $\vect x \geq \vect 0$, and $\vect x$ is a feasible point. The same sets $S$ and $P$ specify the tight inequalities and $\vect x$ is an extreme point of this feasible region. If it is not extreme, then there exists a point $\vect{\hat{x}}$ with corresponding sets of tight inequalities $\hat{S} \supseteq S$ and $\hat{P} \supseteq P$, one of which is a strict superset. 
Let $\vect{\hat {\beta}} = \displaystyle \frac{\hat{\vect x}}{\vect 1' _{N+2}\cdot \hat{\vect x}}$, and
 $\hat{z} = \min[\Lambda \vect{\hat{\beta}}]$.
It can be easily shown that 
 $\hat{\vect{\beta}}$ is a feasible point to the inequalities 
 $\Lambda \vect \beta \geq z  \mathbf{1}_{N+1}$, and 
 $\vect \beta \geq \vect 0$: Indeed, 
  $\Lambda \vect{\hat{\beta}} \geq \min[\Lambda \vect{\hat{\beta}}] = \hat{z}$, 
  and since $\hat{\vect x} \geq \vect 0$, $\vect{\hat{\beta}} \geq \vect 0$.
   It can also be shown that $\vect{\hat{\beta}}$ has tight inequalities in $\hat{S}$ and $\hat{P}$. 
   Indeed $\forall j \in \hat{P}, \hat{\beta}_j = 0$ (thus $\vect{\hat{\beta}}$ has tight inequalities in $\hat{P}$).
$\forall i \in \hat{S},~\hat{\vect x}$ has tight inequalities in $\hat{S}$, thus [$\Lambda \hat{\vect x}]_i = \min[\Lambda \hat{\vect x}] = 1$.
We divide the last equation with $\vect 1'_{N+2} \cdot \hat{\vect x}$ (constant, can get inside the $\min$) and get 
$[\Lambda (\displaystyle \frac{\hat{\vect x}}{\vect 1'_{N+2} \cdot \hat{\vect x}})]_i = \min[\Lambda \displaystyle \frac{\hat{\vect x}}{\vect 1'_{N+2} \cdot \hat{\vect x}} ] $.
But $ \displaystyle \frac{\hat{\vect x}}{\vect 1'_{N+2} \cdot \hat{\vect x}}  = \vect{\hat{\beta}}$, thus $[\Lambda \vect{\hat{\beta}}]_i = \min[\Lambda \vect{\hat{\beta}}]$, so $\vect{\hat{\beta}}$ has tight inequalities in $\hat{S}$. Thus $[\vect{\hat{\beta}}, \hat{z}]$ has a strict superset of tight inequalities to $[\vect \beta;z]$; this is a contradiction with our hypothesis that $[\vect \beta; z ]$ was an extreme point. Thus, the defendability is maximized at the extreme points of the
  polyhedron $\Lambda \vect x \geq \vect{1}_{N+1}$, $\vect x \geq \vect 0$. Given an extreme point $\vect x$, we compute the distribution $\vect \beta = \vect x/ \|\vect x\|$. Thus, a point $\vect x = (x_0, \ldots, x_{N+1})$ corresponds to a defender's strategy $\vect \beta$, after normalizing it.
  \end{proof}
Adding the constant parameter $Nc_a +\epsilon$ to every element of $\Lambda$ does not change the structure of the Nash equilibria but renders $\Lambda$ strictly positive. We thus avoid the problems that arise when the minimum element of the vector is zero (infinite solution).

\subsection{Form of players' strategies in NE}
In this section we show how the NE in mixed strategies for the two players can be computed in polynomial time. 

\subsubsection{Defender's NE strategy}
As we saw in Lemma~\ref{def_lemma}, the best response strategy of the defender is found by
looking at the extreme points of the polyhedron $\Lambda \vect x \geq \vect{1}_{N+1}$, $\vect x \geq \vect 0$. We call the first type ``inequality" constraints and the second type ``positivity" constraints. We have $N+1$ ``inequality"- and $N+2$ ``positivity" constraints.
 Writing down the ``inequality" constraints, we get
\begin{eqnarray*}
  c_d \cdot x_0 + (Nc_a + \epsilon) \|\vect x\| & \geq & 1 \\ c_d \cdot
  (x_0+x_1) +[(N-1)c_a + \epsilon] \|\vect x\| & \geq &1\\ \vdots \\ c_d
  \cdot (x_0+x_1+\ldots+x_N) +\epsilon \|\vect x\| & \geq &1.
\end{eqnarray*}
Our goal is to eliminate nonextreme points that are not selected by a defender in NE, so that we
reduce the number of points we have to check.


\begin{lemma}
Two points $\vect {x_1}$ and $\vect {x_2}$ on the polyhedron, with $\|\vect {x_1}\| = \|\vect {x_2}\|$, correspond to defender NE strategies $\vect {\beta_1}$ and $\vect {\beta_2}$ respectively with $\min [\Lambda \vect {\beta_1}] = \min [\Lambda \vect {\beta_2}]$. 
\label{minL}
\end{lemma}

\begin{proof}
We showed in Lemma~\ref{def_lemma} that the equation that needs to be solved in NE is $\Lambda \vect x \geq \vect 1$, with $\|\vect x\| = 1/z = 1 / \min[\Lambda \vect \beta]$. Thus, if the norm $\|\vect x\|$ is preserved, the optimal attacker cost $\min[\Lambda \vect \beta]$ is also preserved. Hence $\min [\Lambda \vect {\beta_1}] = \min [\Lambda \vect {\beta_2}]$. 
\end{proof}


\begin{lemma}
An extreme point $\vect x$ satisfies at least one tight inequality.
\label{one}
\end{lemma}

\begin{proof}
If none of the inequalities are tight, we scale the vector $\vect x$ down until one inequality becomes tight. The new vector's set of tight inequalities is a strict superset of those of the original vector, thus the point with no tight inequalities is not extreme. 
\end{proof}

\begin{lemma}
If $\|\vect {x_1}\| = \|\vect {x_2}\|$ and $\vect \mu'  \vect {x_1} < \vect \mu' \vect {x_2}$, then $\vect {x_1}$ corresponds to a defender strategy $\vect {\beta_1}$ with a better defendability, i.e., $\theta(\vect {\beta_1}) > \theta(\vect {\beta_2})$.
\label{defendability}
\end{lemma}

\begin{proof}
Since $\|\vect {x_1}\| = \|\vect {x_2}\|$, $\min[\Lambda \vect {\beta_1}] = \min[\Lambda \vect {\beta_2}]$ (Lemma~\ref{minL}). 
Combining the above result with the definition of the defendability we get
\begin{align*}
\theta(\vect {\beta_1}) - \theta(\vect {\beta_2}) & = \min[\Lambda \vect {\beta_1}] -\vect \mu' \vect{\beta_1} - (\min[\Lambda \vect {\beta_2}] -\vect \mu' \vect {\beta_2}) \\
& = \vect \mu' \vect{\beta_2} -\vect \mu' \vect{\beta_1}.
\end{align*}
Since  $\vect \mu' \vect {x_1} < \vect \mu' \vect {x_2}$, the point $\vect {x_1}$ corresponds to a defender strategy $\vect {\beta_1}$ with a smaller false alarm cost, $\vect \mu' \vect{\beta_1} < \vect \mu' \vect{\beta_2}$. Hence  $\theta(\vect {\beta_1}) > \theta(\vect {\beta_2})$. \end{proof}

\begin{lemma}
  An extreme point $\vect x$ corresponding to a defender NE strategy $\vect \beta$ satisfies exactly one contiguous set (of indices) of tight inequalities. 
  \label{block_lemma}
\end{lemma}

\begin{proof}
An extreme point satisfies at least one tight inequality (Lemma \ref{one}). Suppose there are two tight inequalities with indices $s$ and $f$, with $f > s$, and that
there is at least one loose inequality (with index $k$) between $s$ and $f$. Since $k$ is loose, it should be that $x_k \geq \frac{c_a}{c_d}\|\vect x\|>0$ (after subtracting the loose inequality from the previous tight one).
\eat{Subtracting the two tight inequalities $s^{\text{th}}$ and $f^{\text{th}}$ gives 
$\displaystyle \sum_{i = s+1}^f x_i = (f - s) \frac{c_a}{c_d} \|\vect x\|$.} We make the following transformation
\begin{equation}
\hat{x}_i = 
\begin{cases}
	 x_i & \!\!\!\text{for $i \in \{0, \ldots, k-1\} \cup \{k+2, \ldots, N+1\}$}
\\
x_i - \epsilon_1 & \text{for $i= k$}
\\
 x_i + \epsilon_1 & \text{for $i= k+1$},
\end{cases}
\end{equation}
where $\epsilon_1>0$ is small enough so that $\hat{x}_i\geq  \frac{c_a}{c_d}\|\hat{\vect x}\|$.

The transformation preserves the norm ($\|\vect x\| = \|\hat{\vect x}\|$), but $\vect \mu' \vect x > \vect \mu' \hat{\vect x}$. The latter comes from the fact that 
$\vect \mu' (\vect x - \hat{\vect x}) = \mu_k \cdot (x_k-\hat{x}_k) + \mu_{k+1} \cdot (x_{k+1} - \hat{x}_{k+1}) = \mu_k \cdot \epsilon_1 + \mu_{k+1}\cdot (-\epsilon_1) = \epsilon_1 \cdot (\mu_k - \mu_{k+1}) > 0$, since $\vect \mu$ is a strictly decreasing vector (component-wise). Thus, from Lemma~\ref{defendability}, $\hat{\vect x}$ corresponds to a defender strategy with a better defendability. If there are more than one loose inequalities, we can iteratively make the above transformation until the point $\hat{\vect x}$ has a unique contiguous block of tight inequalities, in which case $\hat{\vect x}$ corresponds a defender NE strategy (a defender's unilateral change of strategy would not result in a better defendability).
\end{proof}

Let $s$ and $f$ be the indices of the first and last tight inequalities respectively. 
\begin{lemma}
  An extreme point $\vect x$ that corresponds to a defender NE strategy $\vect \beta$ has zeros before $s$ and after $f+1$, i.e., $$x_i=0, \forall i \in
  \{0,\ldots, s-1\} \cup \{f+2, \ldots, N+1\}.$$
  \label{zeros_lemma}
\end{lemma}

\begin{proof}
We first show that $x_{i}=0, \forall i<s$. If $\exists i \in \{0,\ldots, s-1\}, \text{
s.t. } x_i > 0$, we reduce $x_i$ to $\hat{x}_i$ until either $\hat{x}_i = 0$ or until the $i^\text{th}$ (previously loose) inequality becomes tight, and increase $x_{i+1}$
by the same amount. We maintain $\|\vect x\|$ constant, but we get one more tight constraint. Thus the original point is not extreme, as we 
can find another point whose tight constraints is a strict superset of those of the original. 

 We now show that $x_i=0, \forall i>f+1$. If $\exists
i \in \{f+2, \dots, N+1\}, \text{ s.t. } x_i > 0$, we reduce $ x_i$
until $\hat{x}_i = 0$, and increase $x_{f+1}$
by the same amount. The previously loose $(f+1)^\text{th}$ inequality is made looser, and we keep the norm $\|\vect x\|$ constant, but $\hat{\vect x}$ has one more tight constraint, thus $\vect x$ was not extreme. 
\end{proof}

\begin{lemma}
In any Nash equilibrium, $f = N.$
\label{fN}
\end{lemma}

\begin{proof}
Suppose that $f<N$. Subtracting the tight inequality $f$ from the loose inequality $f+1$, we get $x_{f+1} > x_m$, where $x_m = \displaystyle \frac{c_a}{c_d} \|\vect x\|$. We make the following transformation
\begin{equation}
\hat{x}_i = 
\begin{cases}
	  x_i & \text{for $i \in \{0, \ldots, f\} \cup \{f+3, \ldots, N+1\}$}
\\
x_m  & \text{for $i= f+1$}
\\
x_{f+1} - x_m   & \text{for $i= f+2$.}
\end{cases}
\end{equation}
With the above transformation we get 
\begin{align*}
\vect \mu' (\hat{\vect x} - \vect x)& = \mu_{f+1} \cdot (\hat{x}_{f+1} - x_{f+1}) + \mu_{f+2} \cdot (\hat{x}_{f+2} -  x_{f+2}) \\
					    & = \mu_{f+1} \cdot (x_m - x_{f+1}) + \mu_{f+2} \cdot  (x_{f+1} - x_m - 0) \\
					    & = (x_{f+1} - x_m) \cdot (\mu_{f+2} - \mu_{f+1}) \\
					    & < 0,  \\
					    \end{align*} since $x_{f+2} = 0$, $x_{f+1} > x_m$, and $\vect \mu$ is a strictly decreasing vector ($\mu_{f+2} < \mu_{f+1}$). Hence, for the new point $\hat{\vect x}$, it holds that $\|\hat{\vect x}\| = \| \vect x\|$, but $\vect \mu' \hat{\vect x} < \vect \mu' \vect x$. Thus from Lemma~\ref{defendability} point $\hat{\vect x}$ corresponds to a defender NE strategy with a better defendability. We can continue making the above transformation until $f=N$, when the defender's payoff cannot be maximized by a unilateral deviation. Hence,  in any Nash Equilibrium, $f=N$.
\end{proof}


\begin{lemma}
An extreme point $\vect x$ that corresponds to a defender NE strategy $\vect \beta$ cannot have both $x_s>0$ and $x_{N+1}>0$.
\label{not_both}
\end{lemma}

\begin{proof}
In Lemmata \ref{block_lemma}--\ref{fN}, we proved that an extreme point $\vect x$ that corresponds to a defender NE strategy satisfies a contiguous block of tight inequalities with nonzero components between $s$ through $N+1$. 
We make the following transformation

\begin{equation}
\hat{x}_i = 
\begin{cases}
	 \gamma \cdot x_i & \forall i \in \{0, \ldots, s-1\} \cup \{s+1, \ldots, N\}
\\
0 & \text{for $i= s$}
\\
\gamma \cdot (x_s + x_i)  & \text{for $i= N+1$,}
\end{cases}
\end{equation} with $\gamma = \displaystyle \frac{1}{1-c_d \cdot x_s} $ or, $\gamma = \displaystyle \frac{1}{\|\vect x\| [(N-s)c_a + \epsilon]}$. The definition of $\gamma$ is such that the $s^\text{th}$ inequality is still tight after the transformation, i.e., 
$$\gamma \cdot \left( c_d \cdot 0 + [(N-s)c_a + \epsilon] \| \vect{x}\| \right) = 1.$$
The loose inequalities before $s$ become looser with the scaling with $\gamma$ ($\gamma >1$), whereas the previously tight inequalities $s+1$ through $N$ are still tight. Indeed, after the above transformation, any previously tight inequality with index $k \in \{s+1, \ldots, N \}$ gives
\begin{align*}
&\gamma \left( c_d  \cdot (0 + x_{s+1} + \ldots +x_k) + [(N-s-k)c_a + \epsilon] \| \vect x\| \right)\\
 &=\gamma \left( c_d(x_s + x_{s+1} + \ldots +x_k) + [(N-s-k)c_a + \epsilon] \| \vect x\| \right)- \gamma c_d x_s \\
& \text{ (after subtracting and adding }\gamma \cdot c_d x_s)\\
&= \gamma - \gamma \cdot c_d x_s \text{ (the } (s+1)^\textbf{th } \text{ inequality was tight before)}\\
&=1 \text{ (from the definition of }\gamma).
\end{align*}
With the above transformation, we get an extra tight constraint ($x_s = 0$), thus the previous point was not extreme. 
\end{proof}

\begin{lemma}
For an extreme point $\vect x$ that corresponds to a defender NE strategy $\vect \beta$, we only have two possible combinations\\
1. $(0, \ldots, 0, x_s = 0, x_{s+1} =  \ldots = x_N = x_m,  x_{N+1} \geq 0)$ \\
2. $(0, \ldots, 0, x_s > 0, x_{s+1} =  \ldots = x_N = x_m,  x_{N+1} = 0)$, where $x_m = \displaystyle \frac{c_a}{c_d} \|\vect x\|$.
\label{forms_lemma}
\end{lemma}

\begin{proof}
From Lemma~\ref{block_lemma}, the inequalities $s$ through $N$ are tight. Subtracting any tight inequality $k$, $k \in \{s+1, \ldots, N\}$ from the first tight inequality ($s^\text{th}$), gives $x_k = x_m$, with $x_m = \displaystyle \frac{c_a}{c_d} \|\vect x\| $. Also, from Lemma \ref{zeros_lemma}, $x_i = 0, \forall i \in \{0, \ldots, s-1\}.$ From Lemma \ref{not_both}, the combination $x_s > 0$ and $x_{N+1} > 0$ is not possible, thus the only possible forms of extreme points $x$ that correspond to a defender NE strategies are the ones described above. 
\end{proof}
We can now state and prove our main Theorem. 
\begin{theorem}
\label{theorem}
In any Nash equilibrium the defender's strategy $\vect \beta$ maximizes the defendability (\ref{def_defendability}). A maximizing value of $\vect \beta$ exists amongst one of the two forms in Table~\ref{thm1_table} for some $s$. If there is only one maximizing $\vect \beta$ amongst vectors of the form in Table~\ref{thm1_table}, then the Nash equilibrium is unique. 
  \begin{table}
    \caption{Defender's strategy in NE ($\beta_m = c_a/c_d$)}
    \centering
    \begin{tabular}{|c|c | c  | c | c | c | c | c | c|}
      \hline
      \#&$\ldots$ & $\beta_s$ & $\beta_{s+1}$ & $\ldots $ & $\beta_N$ &
      $\beta_{N+1}$\\
      \hline
      1.&$0$ & $0$ & $\beta_m$ & $\beta_m$ & $\beta_m$ & $1 -(N-s)
      \beta_m $\\
      \hline
     2. &$0$ & $1 -(N-s) \beta_m$ & $\beta_m$ & $\beta_m$ & $\beta_m$ &
      $0$\\
      \hline
    \end{tabular}
    \label{thm1_table}
  \end{table}
  \label{main_table}
\end{theorem} 
\begin{proof}
 The first statement of  the Theorem is a direct consequence of Lemma~\ref{defend_max}. Combining the results of Lemmata ~\ref{def_lemma} and~\ref{forms_lemma}, we can prove the second statement. In the first type, when $x_s = 0$, the point $\vect x$ corresponds to a defender NE strategy $\vect \beta$ with $\| \vect \beta \| = 1$, thus $\beta_{N+1} = 1 - (N-s) \cdot \beta_m$ or $\beta_{N+1} = 0$ and equivalently for the second type, where $x_s > 0$, we have $\beta_s = 1 - (N-s) \cdot \beta_m$, with $\beta_m = c_a/c_d$. Hence, there exists a Nash equilibrium in which the defender's strategy $\vect \beta$ has one of the two forms in Table~\ref{thm1_table}. 
  \end{proof}
In the case that $(N-s) \beta_m = 1$, then both cases give the same NE strategy
$\vect \beta = (0, \ldots, \beta_1 = \ldots = \beta_N = \beta_m, 0)$. Then, $(N-s) c_a  = c_d$ and since $s \geq 0$, the condition that needs to be satisfied is $c_d \leq N c_a$. 
 
If the defendability LP (\ref{def_LP}) produces a unique maximizer $\vect \beta$, then only one of the above cases will maximize the defendability and therefore the Nash equilibrium will be unique. In the case that the solution of (\ref{def_LP}) is not unique, we could prevent ties between extreme points that vary in $s$ by a small perturbation to the problem (e.g., by modifying the function $\mu$ such that it produces a unique maximizer of the defendability).

\subsubsection{Attacker's NE strategy}

In mixed-strategies NE, each player is indifferent among the strategies in their support. Thus, any pure strategy $i$ with $\beta_i = 1$ in the defender's support yields the same (maximum) payoff $\theta$ to the defender in NE. \eat{
According to the Rank Lemma (2.1.4 in \cite{Comb2011}), let $P =\{\vect{x}:A\vect{x} \geq \vect{b},\vect{x} \geq \vect{0}\}$ and let $\vect{x}$ be an extreme point solution of $P$ such that $x_i > 0$ for each $i$. Then any maximal number of linearly independent tight constraints of the form $A_i \vect x = b_i$ for some row $i$ of A equals the number of variables.
}

\eat{Since the defender NE strategies can be found solving only the $s$ through $N$ tight inequalities (rows) and the $s$ through $N$ or $s + 1$ through $N + 1$ positivity constraints (columns), the number of rows and columns is the same and we can reduce the complexity of the game by letting the attacker solve the optimization problem with a square sub matrix $\Lambda_r$.Thus the attacker's strategy in NE is 

\begin{align}
\label{att_NE1}
\vect \alpha' &= [\vect 0 ; \vect{\alpha'_r} ]\\
\label{att_NE2}
\vect \alpha'_r &= (\theta \cdot \vect 1_{f-s-1}' + \vect \mu_r') \cdot \Lambda_r^{-1}. 
\end{align}}

We have proved that in any Nash equilibrium, the defender is playing with a mix of strategies $\vect \beta$ that maximizes defendability. If there exists a unique maximizer $\vect \beta$ of the defendability and $\hat{\theta}$ is the maximum defendability, the attacker's NE strategy $\vect \alpha$ is uniquely derived by the following procedure. First construct a sub matrix of $\Lambda$ that we call $\Lambda_r$, by keeping only the columns that correspond to the support of the defender's strategy ($s$ through $N$, or $s+1$ through $N+1$, or $s+1$ through $N$) and rows that correspond to the support of the attacker's strategy ($s$ through $N$). 
The defender assigns weight to $\{\beta_{\hat{s}}, \ldots, \beta_{N+1}\}$, according to Table \ref{thm1_table}, where $\hat{s}$ optimizes the defendability, The attacker assigns positive weight to $\{\alpha_{\hat{s}}, \ldots, \alpha_N\}$  according to the equations

\begin{align}
\label{att_NE1}
\vect \alpha' &= [\vect 0 ; \vect{\alpha'_r} ]\\
\label{att_NE2}
\vect \alpha'_r &= (\hat{\theta} \cdot \vect 1' + \vect \mu_r') \cdot \Lambda_r^{-1}. 
\end{align}

This procedure gives a unique $\vect \alpha$. This $\vect \alpha$ must be a valid probability distribution (sum to one and have nonnegative elements) for if otherwise, it would contradict Nash's existence theorem.

If there are multiple choices of $\vect{\beta}$ that maximize the defendability (and this can happen either when both cases in Table~\ref{thm1_table} or various selections for $s$ in the same case, give the same maximal defendability), from Nash's existence theorem (and our analysis), at least one of $\vect \beta$ will yield an $\vect \alpha$ with the above procedure that corresponds to a valid probability distribution.

\eat{For the special case that $c_d \leq N \cdot c_a$, the two forms coincide and result in the same NE.}
The complexity to find the NE is polynomial: O($N^2$) to find the defendability, O($N$) for each case, and O($N^2\log N$) to invert matrix $\Lambda_r$. Since we have proved that $\vect \beta$ has a contiguous set of positive elements, we have reduced all the other degrees of freedom and we make only $N+1$ computations for each case to find the optimal $s$. 
For the special case that $c_d \leq N \cdot c_a$, the two forms coincide and result in the same NE.

\section{Numerical Results/Simulations}
\label{sims}

We conducted various experiments for different sets of parameters $N,~
c_a,~c_d,  c_{fa}$, and $p$, assuming that the spammer attacks with Bernoulli
distribution with parameter $\theta_0$. We followed the procedure described above 
to calculate the strategies of both
players at equilibrium. To validate our theoretical results, we used Gambit
software~\cite{Gambit}. 

\begin{figure}[!t] 
  \centering
  \includegraphics[scale=0.95]{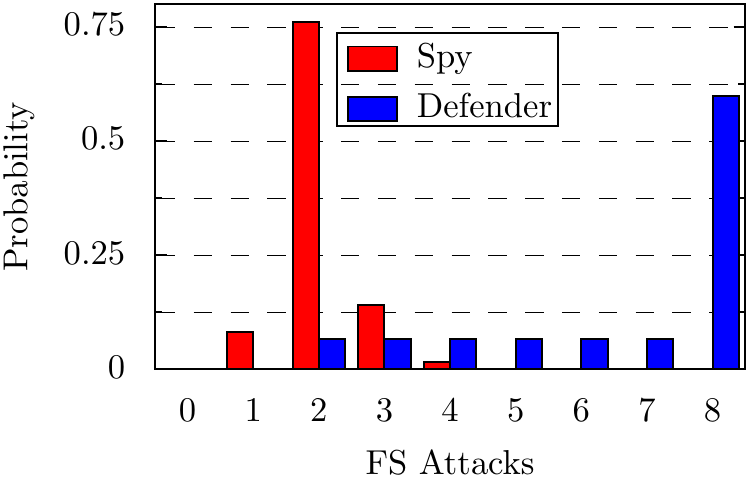}  
  \caption{(Type I) Players's best responses in NE for $N = 7, ~\theta_0 = 0.1,~c_d = 15,~ c_a = 1,~c_{fa}
    = 23, ~p = 0.2.$}
  \label{case1}
\end{figure}

Fig.~\ref{case1} illustrates the first possible type for $N=7$. As we can see, all
the middle points are given the same weight $\beta_m = c_a/c_d= 0.0667$, 
$\beta_s = 0$ and $\beta_{N+1} >\beta_m$. There exists a Nash equilibrium that matches the first row of Table~\ref{thm1_table} with $s=1$.

\begin{figure}[!t] 
  \centering
  \includegraphics[scale=0.95]{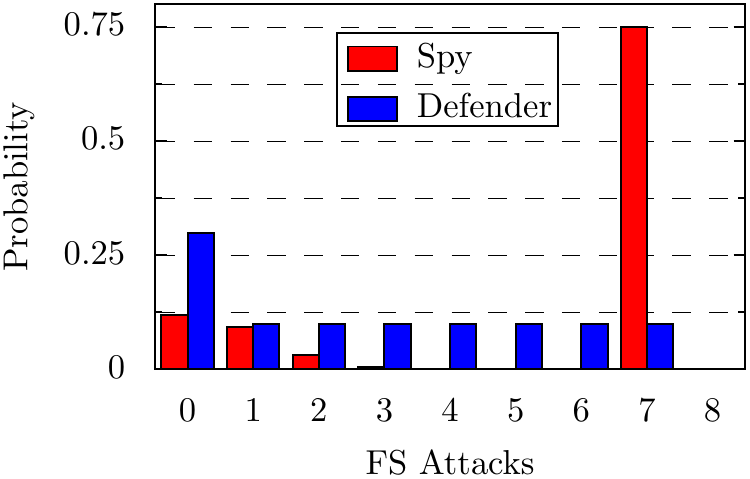}
  \caption{(Type II) Players's best responses in NE for $N = 7, ~\theta_0 = 0.1$, $c_d = 10,~ c_a = 1,~c_{fa}
    = 10, ~p = 0.8$.}
  \label{case2}
\end{figure}

Fig.~\ref{case2} represents the second type of Nash equilibrium strategies for $N=7$. As we can see, all the
middle points are given the same weight $\beta_m = c_a/c_d = 0.1$, but here
$\beta_s > \beta_m$ ($s = 0$) and $\beta_{N+1} = 0$. Note that as $p$ increases, larger weight is given to the smallest
threshold, in order to detect the most-probable-to-exist spy. 

In both figures we observe that the defender gives
positive weight on larger thresholds and is not focused on a range
around $N\theta_0$. \eat{This is explained by the quasi zero-sum nature of the classification problem and the strictly decreasing
false alarm cost function $\mu$: The defender's best response against an attacker strategy, depends only on the defender's choice of $\vect \beta$. Since the false alarm penalty is smaller for larger thresholds, he randomizes among a set of thresholds that include large values to increase his expected payoff.} Every pure strategy (threshold) in the support of the defender's NE strategy must give the defender the same payoff. The attacker's NE strategy $\vect{\alpha}$ is such that he makes the defender's NE payoff for high thresholds the same as for lower ones. This is why the defender gives positive weight to higher thresholds, even when the probability that the spy will attack more than the threshold value is low.

\begin{figure}[t] 
  \centering
  \includegraphics[scale=0.95]{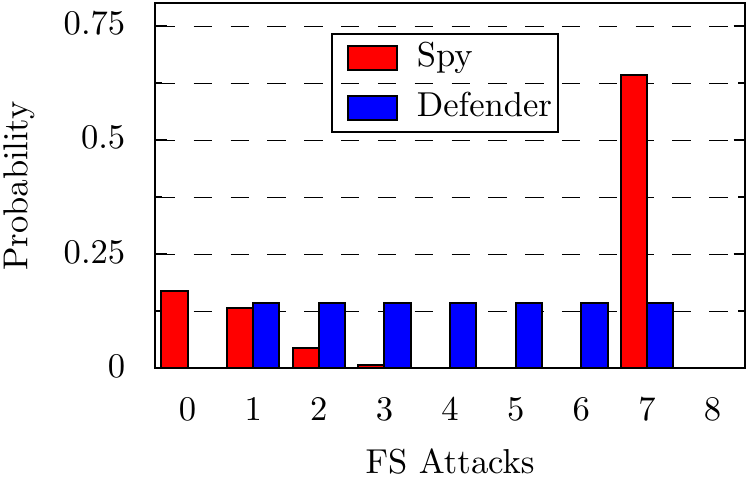}  
  \caption{Type I-II: Defender is uniform when $c_d \leq N\cdot c_a$. Other parameters are $N = 7, ~\theta_0 = 0.1, ~ c_a = 1,~c_{fa}
    = 10, ~p = 0.8$, $c_d = 7$.}
  \label{case0}
\end{figure}

Fig.~\ref{case0} depicts the NE in which both forms coincide, i.e., when $c_d \leq N \cdot c_a$. 
As we can see, the defender NE strategy is uniform and he never classifies the attacker always as spy or always as spammer. 

\begin{figure}[h] 
  \centering
  \includegraphics[scale=0.95]{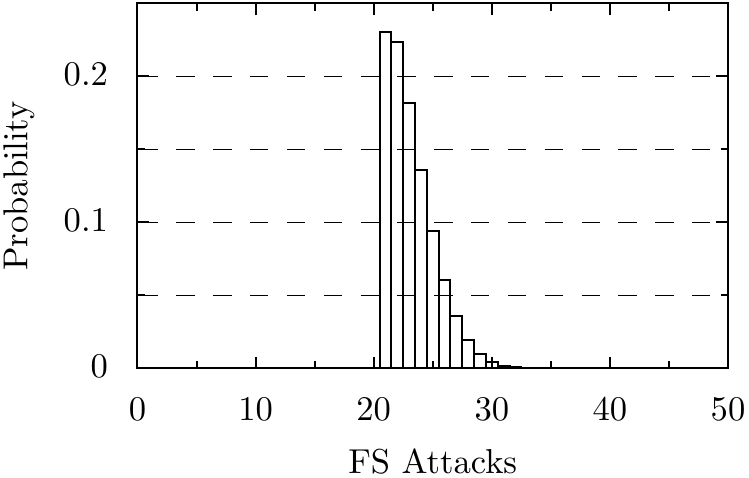}
  \caption{Spy's NE strategy is a truncated version of spammer's distribution. Parameters: $N = 50$, $\theta_0 = 0.4$, $c_d = c_{fa} =142$, $c_a = 1$, $p = 0.3$.}
  \label{truncated}
\end{figure}

 Our simulation results indicated a match between the spy and spammer. In NE, all threshold strategies in the support of the defender give the same payoff. When the defender selects a slightly larger threshold in his support, the decrease in the false alarm cost matches the increase in the misdetection cost, i.e., $\Pr\{H=T\} = \frac{c_{fa}}{c_d} \frac{1-p}{p}\Pr\{Z=T\}$. Hence, the spy's NE strategy is a scaled version of the spammer's distribution. For the spammer strategies that are outside the spy's support in NE, the spy gives zero weight. The spy's NE strategy is a truncated version of the spammer's distribution, as Fig.~\ref{truncated} shows. When either of $p$ or $c_d$ or $c_{fa}$ is large enough, the spy could also put some weight on the ``always attack" strategy --- and so that part of his strategy doesn't look like a truncated spammer distribution.
 
\section{Conclusion}
\label{concl}

We investigated a classification game, where a
network administrator (defender) seeks to classify an attacker as a strategic spy or a naive spammer. We
showed that by taking advantage of the structure of the payoff
formulation, we can characterize and anticipate the structure of the best response strategies of the
two players in polynomial time.  Our experimental results coincide with
the theoretically expected ones: The structure of the cost matrix of the spy
leads to only two forms of defender's strategies in NE. There is a relationship between the spammer's distribution and the spy's NE strategy. Furthermore, the defender NE strategy includes a contiguous set of thresholds that always include large values. If the parameters of the game satisfy a certain condition, the defender is uniformly randomizing among a set of thresholds.

\end{document}